# Coherently driven non-classical light emission from a quantum dot

A. Muller,[1] E. B. Flagg,[1] P. Bianucci,[1] D. G. Deppe,[2] W. Ma,[3] J. Zhang,[3] G. J. Salamo,[3] and C. K. Shih[1*]

[1]Department of Physics, The University of Texas at Austin, Austin, TX 78712

[2]College of Optics and Photonics (CREOL), University of Central Florida, Orlando, FL 32816

[3]Department of Physics, University of Arkansas, Fayetteville, AR 72701

Dated: 07/25/07

**Narrow line-widths and the possibility of enhanced spontaneous emission via coupling to microcavities[1-6] make semiconductor quantum dots ideal for harnessing coherent quantum phenomena at the single photon level[7]. So far, however, all approaches have relied on incoherent pumping[8-11], which limits the desirable properties of the emission. In contrast, coherent excitation was recognized to be necessary for providing both improved photon indistinguishability and high efficiency[8-10], and offers the quantum control capabilities required for basic qubit manipulations[12-14]. Here we achieve, for the first time, resonant and coherent excitation of a quantum dot with simultaneous collection of the non-classical photon emission. Second-order correlation measurements show the unique signature of a coherently-driven two-level quantum emitter[15]: the photon statistics become oscillatory at high driving fields, reflecting the coherent evolution of the excitonic ground state of the quantum dot.**

One of the most elementary features of the light emitted by two-level quantum emitters is photon anti-bunching[15,16], i.e. the tendency of photons to be emitted one by one, rather than in bunches. Observation of this purely non-classical effect in solid-state systems such as semiconductor quantum dots[11,17] (QDs) has aroused a great deal of attention in recent years because it demonstrates their suitability for advanced single photon experiments. QDs additionally benefit from recent advances in micropillar[1-3], microdisk[4,5], and photonic-crystal defect[6] cavity-coupling for efficient light extraction. Therefore, a number of quantum optical and quantum electrodynamic phenomena, familiar from atomic physics, that have wide implications in quantum information processing applications, could in principle be realized in monolithic, scalable solid-state systems.

An important distinction between single-atom and solid-state quantum light sources, however, is that the former are excited resonantly and can therefore be driven *coherently* by a laser[15,16]. This coherent control substantially shapes the non-classical light emission and is also a key capability in successful approaches to deterministic single photon generation with atoms[18,19] and ions[20]. The indirect excitation via continuum or quasi-resonant states in QD-based sources[7] can efficiently pump the emitters, but not coherently. True resonant excitation and collection of the light has not been demonstrated yet, although it is well-known that quantum interference[21] and Rabi oscillations[12-14,22,23] can be achieved in QDs. Extracting the single photons generated by an individual QD that is resonantly driven by strong laser fields has been challenging because of sizable laser scattering in the host crystal that blinds the single photon detection. Thus, so far, QD states have been either probed coherently using non-linear optical techniques without collecting the single photon response[12,13,24], or they have been pumped incoherently via an excited state[8-11].

Based on a new approach to this problem that utilizes a micro-cavity, we show here that a single QD can in fact be coherently driven in resonance fluorescence. The micro-cavity decouples the wave-guided excitation field, introduced in-plane, from the photon emission resonant with a vertical, or Fabry-Perot cavity mode. In the strong-excitation regime fast oscillations coherently induced in the dot by the excitation field profoundly influence the photon statistics. In particular, under continuous wave (CW) excitation, the second-order correlation function, as measured with a high-resolution Hanbury-Brown and Twiss (HBT) setup, exhibits coherent oscillations at the Rabi frequency, markedly different from the basic anti-bunching "dip" observed in non-resonant experiments[15]. These results lay the foundation for a number of very general quantum optical control capabilities which call for simultaneous extraction of light. For example, similarly coherent *pulsed* excitation could form the basis for a deterministic solid-state single photon source.

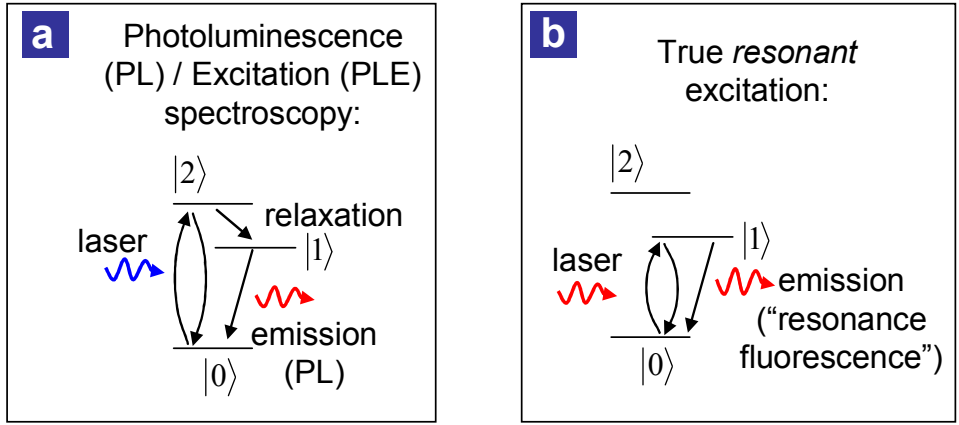

**Figure 1 Energy level diagram for a quantum dot. a,** in photoluminescence spectroscopy. **b,** in resonance fluorescence. State $|0\rangle$ denotes the excitonic vacuum, $|1\rangle$ is the excitonic ground state (one electron-hole pair), and $|2\rangle$ is the first excited state of the dot.

Conventional photoluminescence (PL) or photoluminescence excitation (PLE) spectroscopy (Fig. 1a) naturally relies on fast intra-band carrier relaxation: even when the laser is brought in resonance with an excited state of the dot or with the continuum, emission still primarily occurs from the lowest excitonic state. This is in contrast to true resonant excitation (Fig. 1b), when the laser is exciting the same state from which emission occurs. In order to detect this "resonance fluorescence", self-assembled InGaAs QDs were grown epitaxially between two distributed Bragg reflectors (Fig. 2) making up a two-dimensional micro-cavity that supports both waveguide modes and Fabry-Perot modes. While the sample is maintained at low temperature in a liquid helium cryostat, a single mode optical fibre, mounted on a three-axis inertial walker, is brought within a few microns of the cleaved sample edge. Through this fibre, a tunable CW Ti:Sapphire laser is launched into a waveguide mode of the cavity to excite the dots. The QD emission is spectrally coupled to a Fabry-Perot mode of the cavity and collected by a micro-PL setup equipped with a two-dimensional charge coupled device (CCD) detector mounted on an imaging spectrograph. We focus here on QDs coupled to a cavity mode centered at ~915 nm. When the laser frequency is scanned over the excitonic ground-state of a single QD, the resonance fluorescence is observed as a bright peak in the CCD images, localized both spectrally and spatially. The intensity ratio between the QD emission and residual laser scattering can be as high as 1000:1 due to the orthogonality of the waveguide and Fabry-Perot modes. The laser bandwidth is less than 40 MHz, narrow enough that the total integrated fluorescence intensity as a function of detuning measures the homogeneous line-width of the ground state transition (Fig. 3). For this particular dot we obtain a full width at half maximum (FWHM) of 5.1 μeV ($T_2 = 250 \text{ ps}$) at 7 K.

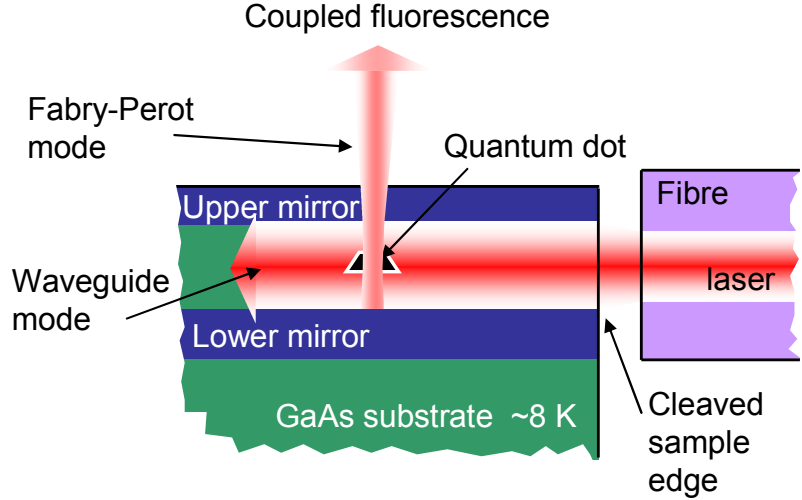

**Figure 2 Experimental setup for orthogonal excitation and detection.** The sample, attached to the cold finger of a liquid He cryostat, is excited from the side with a tunable CW laser that is introduced via a single mode optical fibre. The QD emission coupled to a Fabry-Perot mode of the cavity is efficiently extracted in the orthogonal direction.

The measurements described above demonstrate that the resonant emission from a QD ground state can be extracted essentially background-free while the dot is being coherently driven. Under strong, resonant CW excitation, the Rabi frequency $\Omega = \mu E_0$ resulting from the field-dipole interaction can easily exceed the narrow homogeneous line-width of the dot, given by $1/T_2$. Here $\mu$ denotes the transition dipole moment and $E_0$ is the amplitude of the exciting field. This harmonic driving field (frequency $\omega$) may in general be detuned from the QD transition (resonance frequency $\omega_0$) by an amount $\Delta\omega = \omega - \omega_0$. Throughout, we model the QD as a two-level system composed of a lower state with no exciton, denoted by $|0\rangle$, and an upper state, denoted by $|1\rangle$, in which the dot contains one exciton in its ground state.

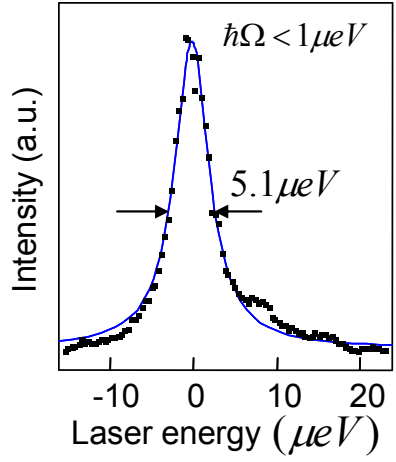

**Figure 3 Spectral lineshape.** The time-integrated emission from a single QD is collected while the exciting laser frequency is scanned over the ground state QD resonance. For this dot we obtain a FWHM of 5.1 µeV.

The evolution of this system follows the optical Bloch equations which treat the excitation field classically and assume a dipole interaction with the QD[25]. The rotating wave approximation simplifies the equations so that the populations of the upper and lower states, $n(t)$ and $m(t)$, and the coherence, $\alpha(t)$, are described by:

$$\frac{d}{dt}n(t) = -i\frac{\Omega}{2}(\alpha(t) - \alpha^*(t)) - \frac{n(t)}{T_1} \quad (1)$$

$$\frac{d}{dt}\alpha(t) = -i\frac{\Omega}{2}(n(t) - m(t)) + i\alpha(t)\Delta\omega - \frac{\alpha(t)}{T_2} \quad (2)$$

where $n(t) + m(t) = 1$ and the phenomenological constants $T_1$ and $T_2$ are respectively the population decay time and dephasing time. These equations have oscillatory solutions, which are the well-known Rabi oscillations, and decay with a time constant $T_2$. Under CW excitation a quasi-steady-state situation is quickly established, and the corresponding expectation values for the population and coherence are obtained from Eq. (1) and (2) as:

$$n_\infty(\Delta\omega) = \frac{1}{2} \frac{\Omega^2 T_1/T_2}{\Delta\omega^2 + T_2^{-2} + \Omega^2 T_1/T_2} \quad (3)$$

$$\alpha_\infty(\Delta\omega) = \frac{i\Omega}{2} \frac{1/T_2 + i\Delta\omega}{\Delta\omega^2 + T_2^{-2} + \Omega^2 T_1/T_2} \quad (4)$$

Since the time-averaged fluorescence intensity is proportional to the population, $n_\infty(\Delta\omega)$, well-known saturation and power-broadening phenomena can be derived from equation (3). On resonance ($\Delta\omega = 0$), the low intensity limit of the line-width equals $2/T_2$, and saturation occurs when the square Rabi frequency, $\Omega^2$, is much greater than the value of $(T_1 T_2)^{-1}$. The experimental data in Fig. 4 show the evolution with Rabi energy and serve as a measure of the excitation intensity (the data provide the proportionality constant between $\Omega^2$ and the laser intensity). The dephasing time, $T_2$, determined from the line-width is 250 ps.

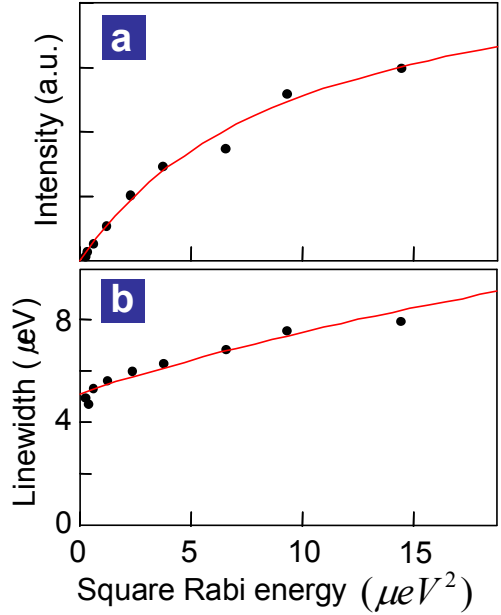

**Figure 4 Steady-state properties. a,** Saturation of the time-integrated emission occurs as the system enters the strong excitation regime. **b,** Power broadening of the spectral lineshape. The excitation linewidth, extracted from lineshapes such as the one shown in Fig. 2, is plotted as a function of square Rabi Energy.

Despite the steady-state excitation and emission intensity, the single QD continues to dynamically evolve due to the excitation field. The information of this evolution, however, is contained in the time-dependent characteristics of the emission and requires the measurement of correlations. The first-order correlation function, for instance, whose Fourier transform is the power spectrum, has been described by Mollow[25] and leads to a well-known spectral triplet. This "Mollow triplet" reflects the coherent population oscillations at the Rabi frequency $\Omega$ that the two-level system undergoes under sufficiently strong driving fields, leading to sidebands at frequencies $\omega \pm \Omega$.

The second-order correlation function, which we measure here, is particularly important in the context of single photon emission. Motivated by the realization of a high-efficiency, high-speed single photon source based on a QD, second-order correlation measurements have been reported in a number of contexts in the past few years. These are typically performed with a HBT setup[11,17] in which two fast photo-detectors and time-counting electronics measure the arrival time difference between photons emerging from a 50/50 beam splitter. If the emission originates from a single quantum emitter, the single photon can only take one of the two beam splitter ports and thus there will be no coincidence counts at zero time delay. This photon "anti-bunching" stands as an important benchmark for single photon sources.

Despite interest in *coherent* quantum optical devices, all QD anti-bunching measurements reported to date have relied on non-resonant, *incoherent* excitation of the dots. Observation of an anti-bunching dip does not, in fact, require coherence: it can be recorded at room temperature[26] and under completely incoherent excitation, such as electrical injection[27]. CW second-order correlations from a *coherently*-driven single emitter, however, reveal a distinct oscillatory signature at high intensities, in addition to photon anti-bunching. The second order correlation function can be obtained from $g^{(2)}(t,\tau) = \langle b^+(\tau) b^+(t+\tau) b(t+\tau) b(\tau) \rangle$, where $b$ and $b^+$ are the field operators. These are proportional to the QD dipole operators $|0\rangle\langle 1|$ and $|1\rangle\langle 0|$, respectively[25], and we obtain ($\Delta\omega = 0$)[28]:

$$g^2(t) = 1 - ae^{-\frac{1}{2}(\frac{1}{T_1}+\frac{1}{T_2})} \{\cos(\Omega' t) + \frac{(1/T_1 + 1/T_2)/2}{\Omega'} \sin(\Omega' t)\} \qquad (5)$$

where $\Omega' = \sqrt{\Omega^2 - \frac{1}{4}(\frac{1}{T_1} - \frac{1}{T_2})^2}$ and $a$ accounts for a constant background correction. When $\Omega << 1/T_2$, Eq. (5) reduces to the familiar anti-bunching dip, whose width, however, is now determined by both $T_1$ and $T_2$. This is unlike the incoherent situation, where the width of the notch is usually assumed to be given only by $T_1$. When $\Omega \gg 1/T_2$, on the other hand, there are oscillations that persist within a time $T_2$. These oscillations originate in the coherent population oscillations that the dot undergoes, and are the same that give rise to the Mollow triplet[25,28]. Since the field creation and annihilation operators are proportional to the atomic operators, the dot's population oscillations result in a photon state that in general is in a superposition of vacuum and single photon states. Upon detection of a "start" photon at one of the beam splitter outputs in the HBT setup, the system is guaranteed to be in the ground state. Therefore, the arrival probability of a "stop" photon at the other output measures the amplitude in the superposition state of a photon that had evolved coherently with the field for a time $t$, the delay between the two detections. Mathematically, this can be understood from a factorization of the second-order correlation function into two parts[16,29]: the mean steady-state intensity from the QD, $n_\infty$, and the mean time-dependent intensity, $n(t)$, from a QD that started in the ground state at $t = 0$, due to a quantum "jump" into the ground state immediately following photon emission.

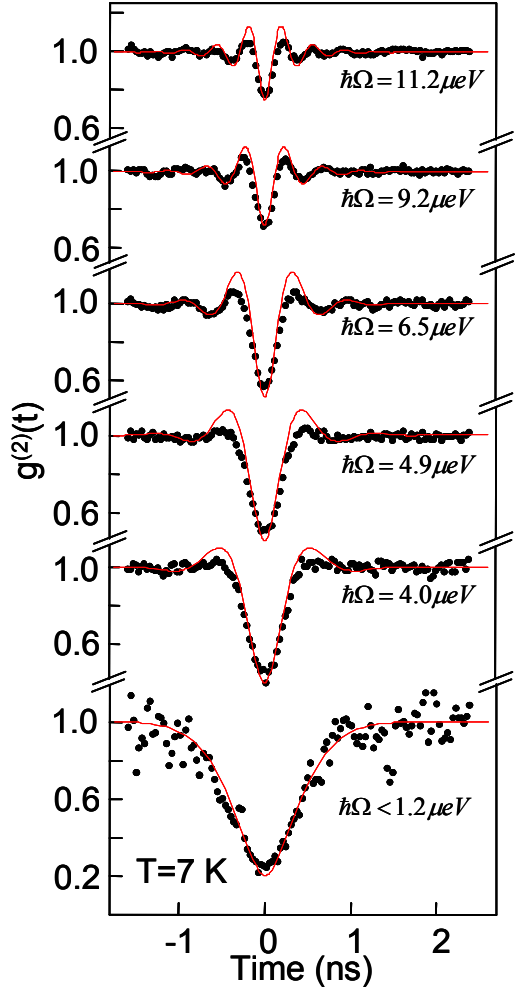

**Figure 5 Second-order correlation function.** $g^{(2)}(t)$ is recorded with a Hanbury-Brown and Twiss setup for various excitation amplitudes.

We observe these exact features in our experimental measurements. A high-resolution HBT setup (~80 ps time resolution) is used to measure the second order correlation function of the light emitted by the same dot as in Fig. 3 and Fig. 4. In Fig. 5 it is plotted for various excitation amplitudes, increasing from bottom to top. The anti-bunching evolves into the oscillatory function described by Eq. (5), with a frequency increasing with excitation amplitude. At the highest intensity, several periods of oscillations are clearly visible. In these measurements (~30 min integration time), no spectral filter was used, but a confocal aperture in the collection path spatially filtered the emission from nearby dots. The residual laser scattering, which by itself gives rise to a constant second order correlation function, was subtracted from the raw data. At the lowest (highest) intensity shown, this contribution was 5% (30%). The corrected data were then fit using Eq. (5) and the extracted Rabi energies are displayed in Fig. 5. The decay times used for the fit were $T_1 = 350\,\text{ps}$ and $T_2 = 250\,\text{ps}$. Better quantitative assessment will require separate direct measurements of $T_1$, for example using time resolved single photon counting. Nonetheless, $T_2 < 2T_1$ is frequently reported from QDs, and might be due to elastic contributions from exciton-phonon scattering or to measurement inhomogeneities, such as spectral diffusion, that occur on time scales faster than our acquisition times (~1 s). If loss of coherence was only due to radiative processes we would expect $T_2 = 2T_1$.

The possibility of coherently controlling the photon emission from a single solid-state emitter offers new opportunities for quantum information applications. For example, almost all proposed schemes for implementing basic qubit-photon interfaces strictly require resonant coherent control with simultaneous extraction of the photons. Together with cavity quantum electrodynamics, coherent control is also necessary for deterministic single photon sources with high efficiency and indistinguishability[8], based on recent results with trapped atoms and ions[18-20]. Such realization will necessitate extending the techniques presented here to pulsed excitation and QDs in three-dimensional micro-cavities, in which the cavity can substantially modify the dots' emission properties. This could be achieved with the all-epitaxial micro-cavities described in Ref. [30] which are ideally suited to introduce a wave-guided laser, but also with photonic crystal defect micro-cavities, or microdisks using the correct geometry.

In conclusion, we have demonstrated coherently controlled non-classical light emission from a single QD in a planar micro-cavity. The cavity is used to extract the light emitted at the same frequency as the strong excitation laser, nearly background-free. Consequently, the second-order correlation function of this single QD resonance fluorescence could be measured both in the weak and strong excitation regime. When the Rabi frequency exceeds the decoherence rate of the dot, the coherent oscillations are directly visible in the second-order correlation function as high frequency oscillations. These results are a first step towards a coherently driven single photon source with both high efficiency and indistinguishability.

**Acknowledgements**

The authors acknowledge financial support from the National Science Foundation (DMR-0210383 and DMR-0606485 and DGE-054917), and the W. Keck foundation.